\documentclass[english,aps,manuscript]{revtex4}
\usepackage[T1]{fontenc}
\usepackage[latin9]{inputenc}
\usepackage{textcomp}
\usepackage{amsthm}
\usepackage{amsmath}
\usepackage{graphicx}
\usepackage{amssymb}
\usepackage{esint}

\makeatletter

\newcommand{\lyxmathsym}[1]{\ifmmode\begingroup\def\b@ld{bold}
  \text{\ifx\math@version\b@ld\bfseries\fi#1}\endgroup\else#1\fi}

\@ifundefined{textcolor}{}
{%
 \definecolor{BLACK}{gray}{0}
 \definecolor{WHITE}{gray}{1}
 \definecolor{RED}{rgb}{1,0,0}
 \definecolor{GREEN}{rgb}{0,1,0}
 \definecolor{BLUE}{rgb}{0,0,1}
 \definecolor{CYAN}{cmyk}{1,0,0,0}
 \definecolor{MAGENTA}{cmyk}{0,1,0,0}
 \definecolor{YELLOW}{cmyk}{0,0,1,0}
 }
\numberwithin{equation}{section}
  \theoremstyle{plain}
  \newtheorem*{thm*}{\protect\theoremname}

\makeatother

\usepackage{babel}
  \providecommand{\theoremname}{Theorem}

\begin{document}

\title{$H-$theorems for the Brownian motion on the hyperbolic plane}

\author{C. Vignat and P. W. Lamberti}
\address{L.S.S., Supelec, France and Facultad de 
Matematica,
Astronomia y Fisica, Universidad Nacional de Cordoba and CONICET, Argentina\,\,}
\email{vignat@u-psud.fr}

\begin{abstract}
We study $H-$theorems associated with the Brownian motion with constant
drift on the hyperbolic plane. Since this random process verifies
a linear Fokker-Planck equation, it is easy to show that, up to a
proper scaling, its Shannon entropy is increasing over time. As a
consequence, its distribution is converging to a maximum Shannon entropy
distribution which is also shown to be related to the non-extensive
statistics. In a second part, relying on a theorem by Shiino, we extend
this result to the case of Tsallis entropies: we show that under a
variance-like constraint, the Tsallis entropy of the Brownian motion
on the hyperbolic plane is increasing provided that the non-extensivity
parameter of this entropy is properly chosen in terms of the drift
of the Brownian motion.
\end{abstract}
\maketitle

\section{Introduction}

The existence of an $H-$theorem for a statistical system submitted
to some constraints ensures that the Shannon entropy $h\left(f_{t}\right)$
of its probability density function (p.d.f.) $f_{t}$ at time $t,$
\[
h\left(f_{t}\right)=-\int_{\mathbb{R}}f_{t}\left(x\right)\log f_{t}\left(x\right)dx,
\]
is increasing with time. As a consequence, the asymptotic (stationary)
p.d.f. $f_{\infty}\left(x\right)$ of the system is the maximum Shannon
entropy p.d.f. that satisfies the given constraints.

The Shannon entropy is a particular member of a family of information
measures called Tsallis entropies that were introduced \cite{Tsallis-1}
in 1988 by Tsallis i the context of statistical physics; they are
defined as\[
h_{q}\left(f_{t}\right)=\frac{1}{1-q}\left(\int_{\mathbb{R}}f_{t}^{q}\left(x\right)dx-1\right);
\]
where $q>0$ is the non-extensivity parameter. It can be checked using
l'Hospital rule that the Shannon entropy is the limit case $\lim_{q\to1}h_{q}\left(f_{t}\right)=h\left(f_{t}\right).$

A natural question then arises: under what conditions does an $H-$theorem
extend to an $H_{q}-$theorem, where the Shannone entropy is replaced
by a Tsallis entropy ? Several studies have been devoted to this problem
in the recent years: for example, Plastino et al \cite{Plastino}
study the conditions of existence of an $H_{q}-$theorem for the following
non-linear Fokker-Planck equation\[
\frac{\partial f_{t}}{\partial t}=-\frac{\partial}{\partial x}\left(K\left(x\right)f_{t}\left(x\right)\right)+\frac{1}{2}Q\frac{\partial^{2}}{\partial x^{2}}f_{t}^{2-q}\left(x\right)
\]
for some parameter $q,$ while Tsallis and Bukman \cite{Tsallis}
apply the same approach to the equation \[
\frac{\partial f_{t}^{\mu}\left(x\right)}{\partial t}=-\frac{\partial}{\partial x}\left(F\left(x\right)f_{t}^{\mu}\left(x\right)\right)+D\frac{\partial^{2}}{\partial x^{2}}f_{t}^{\nu}\left(x\right)
\]
for some positive parameters $\mu$ and $\nu.$

Our aim in this paper is to show the existence of both an $H-$ and
an $H_{q}-$theorem for the Brownian motion - more precisely its $x-$component
- with constant drift on the hyperbolic plane. This study is simplified
by the fact that this component of the Brownian motion satisfies a
linear Fokker-Planck equation, for which the conditions of existence
of $H-$theorems are well-known, as we will see below. However, even
in this simple case, this study reveals an interesting link between
the constant drift parameter of the Brownian motion, the constant
negative curvature of the hyperbolic plane and the non-extensivity
parameter $q$ that characterizes the entropy.

\section{The classical entropic approach to the linear Fokker-Planck equation}

\subsection{General approach}

In the case of a system described by a univariate p.d.f. $f_{t}\left(x\right)$
that verifies the linear Fokker-Planck equation\begin{equation}
\frac{\partial f_{t}\left(x\right)}{\partial t}=-\frac{\partial}{\partial x}\left(K\left(x\right)f_{t}\left(x\right)\right)+\frac{\partial^{2}}{\partial x^{2}}\left(Q\left(x\right)f_{t}\left(x\right)\right),\label{eq:FPequation}
\end{equation}
where $K\left(x\right)$ is the drift function and $Q\left(x\right)$
the diffusion function, it can be shown (see for example \cite{Risken})
that the relative entropy (or Kullback-Leibler divergence)\[
h\left(f_{t}\Vert g_{t}\right)=\int_{\mathbb{R}}f_{t}\left(x\right)\log\frac{f_{t}\left(x\right)}{g_{t}\left(x\right)}dx
\]
between two any solutions of (\ref{eq:FPequation}) decreases to $0$
with time. More precisely, it holds\begin{equation}
\frac{\partial}{\partial t}h\left(f_{t}\Vert g_{t}\right)=-\int_{\mathbb{R}}Q\left(x\right)f_{t}\left(x\right)\left(\frac{\partial}{\partial x}\log\frac{f_{t}\left(x\right)}{g_{t}\left(x\right)}\right)^{2}dx\le0\label{eq:derivativeKL}
\end{equation}
since the diffusion function is assumed positive. Thus entropy is
decreasing and bounded, so that the limit distributions $f_{\infty}$
and $g_{\infty}$ verify\[
\int_{\mathbb{R}}Q\left(x\right)f_{t}\left(x\right)\left(\frac{\partial}{\partial x}\log\frac{f_{t}\left(x\right)}{g_{t}\left(x\right)}\right)^{2}dx=0,
\]
which implies that $f_{\infty}$ and $g_{\infty}$ coincide. This
proves the unicity of a stationary solution of the Fokker-Planck equation
(\ref{eq:FPequation}).

We note that the quantity\begin{equation}
I\left(f\Vert g\right)=\int_{\mathbb{R}}f\left(\frac{\partial}{\partial x}\log\frac{f\left(x\right)}{g\left(x\right)}\right)^{2}dx\label{eq:Fisher}
\end{equation}
is nothing but the relative Fisher information \cite[eq. (174)]{Villani}
between the p.d.f.s $f$ and $g;$ thus the integral that appears
on the right-hand side of (\ref{eq:derivativeKL}) is a weighted version
of this relative Fisher information, with the diffusion function $Q\left(x\right)$
as the weighting function.

In order to deduce an $H-$theorem from this result, we denote as
$g_{\infty}$ the stationary solution to (\ref{eq:FPequation}), assuming
that it exists. We then remark that the relative entropy between any
solution $f_{t}$ of (\ref{eq:FPequation}) and the stationary solution
$g_{\infty}$ is related to the Shannon entropy of $f_{t}$ as\[
h\left(f_{t}\Vert g_{\infty}\right)=-h\left(f_{t}\right)-\int_{\mathbb{R}}f_{t}\left(x\right)\log g_{\infty}\left(x\right)dx.
\]
Thus, provided that the solution $f_{t}$ verifies \textit{at any
time} the constraint\begin{equation}
\int_{\mathbb{R}}f_{t}\left(x\right)\log g_{\infty}\left(x\right)dx=\eta_{1}\label{eq:general constraint}
\end{equation}
where $\eta_{1}$ is a constant, we deduce that as the relative entropy
decreases to $0,$ the Shannon entropy of the solution $f_{t}$ increases
with time to its maximum value. 

We remark that, in a statistical physics framework, the relative entropy
$h\left(f_{t}\Vert g_{\infty}\right)$ coincides with the free energy.
Moreover, taking the limit $t\to+\infty$ in (\ref{eq:general constraint})
shows that the constraint $\eta_{1}$ is also equal to the negentropy
of the stationary solution $g_{\infty},$ namely\begin{equation}
\int_{\mathbb{R}}g_{\infty}\left(x\right)\log g_{\infty}\left(x\right)dx=-h\left(g_{\infty}\right).\label{eq:negentropy}
\end{equation}

\subsection{The $x-$component of the Brownian motion in the Poincaré half-upper
plane}

In \cite{Comtet}, Comtet et al. derived the differential equation
verified by the $x-$component $X_{t}$ of the Brownian motion with
constant drift $\mu$ and constant diffusion constant $D$ in the
Poincaré half-upper plane representation of the hyperbolic plane,\[
\frac{\partial}{\partial t}f_{t}\left(x\right)=D\frac{\partial}{\partial x}\left[\left(1+x^{2}\right)\frac{\partial}{\partial x}f\left(x\right)+\left(2\mu+1\right)xf_{t}\left(x\right)\right].
\]
This is a linear Fokker-Planck equation; in the notations of (\ref{eq:FPequation}),
the diffusion function is quadratic and positive, $Q\left(x\right)=D\left(1+x^{2}\right)$
and the drift function is linear, $K\left(x\right)=D\left(1-2\mu\right)x.$
Moreover, for a positive drift $\mu,$ the asymptotic solution reads\begin{equation}
g_{\infty}\left(x\right)=A_{\mu}\left(1+x^{2}\right)^{-\mu-\frac{1}{2}}\label{eq:asymptotic}
\end{equation}
with a normalization constant $A_{\mu}=\frac{\Gamma\left(\mu+\frac{1}{2}\right)}{\Gamma\left(\mu\right)\Gamma\left(\frac{1}{2}\right)}.$
The constraint to be verified by the p.d.f. $f_{t}$ is thus deduced
from the above results as\begin{equation}
\int_{\mathbb{R}}f_{t}\left(x\right)\log\left(1+x^{2}\right)dx=\eta_{1}\,\,\forall t>0.\label{eq:log constraint}
\end{equation}
(note that the normalization constant $\log A_{\mu}$ and the constant
$-\mu-\frac{1}{2}$ that appear in $\log g_{\infty}$ need not be
taken into account). 

The value of the constant $\eta_{1}$ can be easily computed as\[
\eta_{1}=\psi\left(\mu+\frac{1}{2}\right)-\psi\left(\mu\right)
\]
where $\psi$ is the digamma function.

We note that this constraint can be imposed by a simple scaling of
the random process $X_{t}$ since the function\[
a\mapsto\int_{\mathbb{R}}f_{t}\left(x\right)\log\left(1+ax^{2}\right)dx
\]
is a bijection from $\mathbb{R}^{+}$ to $\mathbb{R}^{+}.$ We also
remark that this scaling does not require the existence of a variance
for $X_{t};$ for example, the Cauchy p.d.f.\[
f_{C}\left(x\right)=\frac{1}{\pi}\frac{1}{1+x^{2}},\,\, x\in\mathbb{R}
\]
- which is the asymptotic distribution of the Brownian motion without
drift $\left(\mu=0\right)$ on the hyperbolic plane - has an infinite
variance but finite nonlinear moment\[
\int_{\mathbb{R}}f_{C}\left(x\right)\log\left(1+x^{2}\right)dx=2\log2.
\]

We deduce the following theorem.
\begin{thm*}
The Shannon entropy of the $x-$component, normalized according to
(\ref{eq:log constraint}), of the Brownian motion on the Poincaré
half-upper plane representation of the hyperbolic plane increases
over time.
\end{thm*}

\section{A generalization to the Tsallis entropies}

\subsection{General approach}

The monotone behavior of the relative Shannon entropy between any
two solutions $f_{t}$ and $g_{t}$ of the linear Fokker-Planck equation
(\ref{eq:FPequation}) has been extended by Shiino \cite{Shiino}
to the case of the relative Tsallis entropy, defined as\[
h_{q}\left(f_{t}\Vert g_{t}\right)=\frac{1}{q-1}\left(\int_{\mathbb{R}}f^{q}\left(x\right)g^{1-q}\left(x\right)dx-1\right).
\]
More precisely, the derivative with respect to time of this relative
entropy verifies %
\footnote{The proof of this result is omitted in \cite{Shiino}; we provide
it in the annex for the interested reader.%
}\begin{equation}
\frac{\partial}{\partial t}h_{q}\left(f_{t}\Vert g_{t}\right)=-q\int_{\mathbb{R}}D\left(x\right)f_{t}\left(x\right)\left(\frac{f_{t}\left(x\right)}{g_{t}\left(x\right)}\right)^{q}\left(\frac{\partial}{\partial x}\log\frac{f_{t}\left(x\right)}{g_{t}\left(x\right)}\right)^{2}dx\le0,\,\,\forall q>0.\label{eq:derivativeq}
\end{equation}
We note that this inequality holds for any positive value of $q$
and simplifies to (\ref{eq:derivativeKL}) as $q\to1.$ Moreover,
the right-hand side integral in (\ref{eq:derivativeq}) can be considered
as a $q-$version of the relative Fisher information that appears
in (\ref{eq:derivativeKL}).

In order to deduce from this monotonicity an $H_{q}-$theorem, we
need the additional assumption that the stationary solution $g_{\infty}$
can be written under the form\[
g_{\infty}\left(x\right)=C_{q}\left(1+U\left(x\right)\right)^{\frac{1}{1-q_{*}}}
\]
for some specific value $q=q_{*}$ of the non-extensivity parameter.
This assumption means that the p.d.f. $g_{\infty}$ is itself a maximum
$q_{*}-$entropy pdf with constraint\begin{equation}
\int g_{\infty}^{q_{*}}\left(x\right)U\left(x\right)dx=\eta_{q_{*}}\label{eq:qconstraint}
\end{equation}
for some constant value $\eta_{q_{*}}.$ We refer the reader to \cite{Borland}
for the conditions on the diffusion and drift functions of the Fokker-Planck
equation that ensure the validity of this assumption. In this case,
we have\[
h_{q_{*}}\left(f_{t}\Vert g_{\infty}\right)=-C_{q_{*}}^{1-q_{*}}h_{q_{*}}\left(f_{t}\right)-\frac{C_{q_{*}}^{1-q_{*}}}{1-q_{*}}\int f_{t}^{q_{*}}\left(x\right)U\left(x\right)dx-\beta_{q_{*}}
\]
where $\beta_{q_{*}}=-\frac{A_{q_{*}}^{1-q_{*}}}{1-q_{*}}$ is a constant.
Thus, provided that the constraint\[
\int_{\mathbb{R}}f_{t}^{q_{*}}\left(x\right)U\left(x\right)dx=\eta_{q_{*}}
\]
is met \textit{at all times}, we deduce that the $q_{*}-$entropy
is increasing with time and reaches asymptotically the entropy of
the stationary pdf $g_{\infty}.$

\subsection{The $x-$component of the Brownian motion in the Poincaré half-upper
plane}

In the special case of the $x-$component $X_{t}$ of the Brownian
motion in the Poincaré half-upper plane, the stationary solution (\ref{eq:asymptotic})
verifies the condition (\ref{eq:qconstraint}) with the value $q_{*}$
such that $-\mu-\frac{1}{2}=\frac{1}{1-q_{*}}$ and the function $U\left(x\right)=x^{2}$,
namely

\begin{equation}
\frac{1}{1-q}\int_{\mathbb{R}}x^{2}f_{t}^{q_{*}}\left(x\right)dx=\eta_{q_{*}}\label{eq:q-constraint}
\end{equation}
with \begin{equation}
q_{*}=\frac{2\mu+3}{2\mu+1},\,\,1<q_{*}<3.\label{eq:q*}
\end{equation}
As a consequence, since the $q-$relative entropy is decreasing, the
$q-$entropy is increasing under the conditions that $q$ is chosen
equal to $q_{*}$ as in (\ref{eq:q*}) and that the constraint (\ref{eq:q-constraint})
is verified. 

Moreover, by taking the limit as $t\to+\infty$ in (\ref{eq:q-constraint}),
the constraint $\eta_{q_{*}}$ in (\ref{eq:q-constraint}) is computed
as \[
\eta_{q_{*}}=\frac{1}{1-q}\int_{\mathbb{R}}x^{2}g_{\infty}^{q_{*}}\left(x\right)dx.
\]
 The value of this constraint is\[
\eta_{q_{*}}=\pi^{\frac{q_{*}}{2}}\frac{\left(q_{*}-1\right)}{2}\left(\frac{\Gamma\left(\frac{q_{*}-3}{2q_{*}-2}\right)}{\Gamma\left(\frac{1}{q_{*}-1}\right)}\right)^{q_{*}}.
\]
We note that this constraint can always be imposed by scaling: denoting
$\sigma_{q}^{2}\left(X_{t}\right)=\left(1-q\right)^{-1}\int x^{2}f_{t}^{q}\left(x\right)dx$
the value of this constraint and $\sigma_{q}^{2}\left(Y\right)$ the
one corresponding to $Y=aX$ with $a>0,$ then\[
\sigma_{q}^{2}\left(Y\right)=a^{3-q}\sigma_{q}^{2}\left(X\right).
\]

We deduce the following result.
\begin{thm*}
With $q=q_{*}$ as in (\ref{eq:q*}), the $q-$entropy of the Brownian
motion, normalized according to (\ref{eq:q-constraint}), is increasing
with time. 
\end{thm*}

\section{Numerical Illustration}

The following figures depict ten realizations of a discretized version
of this Brownian motion, with parameters $D=0.01,$ $m=0$ and $\mu=1$
on Figure \ref{fig:figure1} while $D=0.01,$ $m=3$ and $\mu=7$
on Figure. In both cases, the process starts from the point $x=0,\,\, y=1$
and the superimposed thick curve is the asymptotic probability density
of its $x\lyxmathsym{\textminus}$component. Without external drift
(Figure \ref{fig:figure1}), the random process wanders for a long
time far away from the real axis before \textquotedblright{}falling\textquotedblright{}
on it. The distribution of the \textquotedblright{}landing points\textquotedblright{}
on the real axis is thus very wide, in fact a Lorentz distribution
with infinite variance. With an external drift (Figure \ref{fig:figure2}),
the process is forced to walk in the direction of the real axis, so
that the landing points are more concentrated around $0$, what is
reflected by their narrow distribution, a $q\lyxmathsym{\textminus}$Gaussian
distribution with variance $\sigma^{2}=0.2.$

\begin{center}
\begin{figure}[h]
\begin{centering}
\includegraphics[scale=0.5]{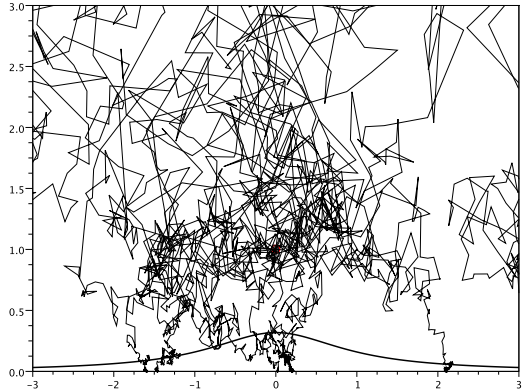}
\par\end{centering}

\caption{\label{fig:figure1}ten realizations of a discretized version of the
Brownian motion on the hyperbolic plane, with parameters $D=0.01,$
$m=0$ and $\mu=1$}

\end{figure}
\begin{figure}[h]
\begin{centering}
\includegraphics[scale=0.5]{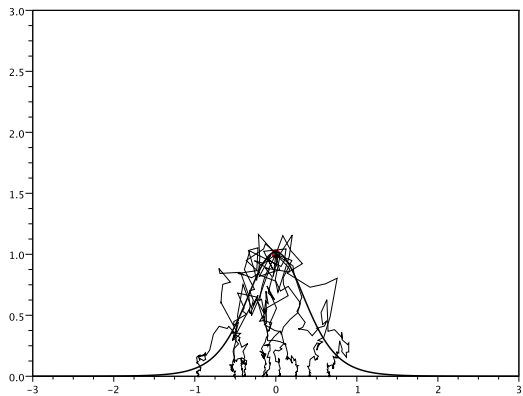}
\par\end{centering}

\caption{\label{fig:figure2}ten realizations of a discretized version of the
Brownian motion on the hyperbolic plane, with parameters $D=0.01$,
$m=3$ and $\mu=7$}

\end{figure}

\par\end{center}

\section{The Brownian motion in the unit disk}

Another representation of the hyperbolic space is the unit disk $\mathbb{D}=\left\{ w=re^{i\theta},r\le1,0\le\theta<2\pi\right\} $
with the metric in polar coordinates\[
ds^{2}=\frac{4}{\left(1-r^{2}\right)^{2}}\left(dr^{2}+r^{2}d\theta^{2}\right).
\]

There is a conformal mapping between the Poincaré upper half-plane
$\mathbb{H}=\left\{ z=x+iy,\, y>0\right\} $ and the unit disk $\mathbb{D}$
defined as\[
w=\frac{iz+1}{z+i}.
\]
Comtet et al show that the density of the radial component of the
Brownian motion in the unit disk representation of the hyperbolic
space converges to $\delta\left(r-1\right)$ as $t\to+\infty$. Having
no explicit Fokker-Planck for the radial part $\theta_{t}$ of this
process, we were unable to prove a corresponding $H-$theorem. However,
we show here that we can use the maximum entropy approach to derive
the asymptotic distribution of the angular part $\theta_{t}$, using
the following result:
\begin{thm*}
If the random variable $X$ has maximum entropy under the log-constraint
$E\log\left(1+X^{2}\right)=\gamma$, then the random variable\[
\tilde{X}=\frac{X}{\sqrt{1+X^{2}}}
\]
has maximum entropy under the constraint $E\log\left(1-\tilde{X}^{2}\right)=-\gamma$. 

More precisely, if the p.d.f. of $X$ reads \[
f_{X}\left(x\right)=\frac{\Gamma\left(\mu+\frac{1}{2}\right)}{\Gamma\left(\frac{1}{2}\right)\Gamma\left(\mu\right)}\left(1+x^{2}\right)^{-\mu-\frac{1}{2}},\,\, x\in\mathbb{R}
\]
then $E\log\left(1+X^{2}\right)=\psi\left(\mu+\frac{1}{2}\right)-\psi\left(\mu\right)$
and the p.d.f. of $\tilde{X}$ reads\[
f_{\tilde{X}}\left(\tilde{x}\right)=\frac{\Gamma\left(\mu+\frac{1}{2}\right)}{\Gamma\left(\frac{1}{2}\right)\Gamma\left(\mu\right)}\left(1-\tilde{x}{}^{2}\right)^{\mu-1},\,\,-1\le\tilde{x}\le+1
\]
with $E\log\left(1-\tilde{X}^{2}\right)=-E\log\left(1+X^{2}\right)=\psi\left(\mu\right)-\psi\left(\mu+\frac{1}{2}\right).$
\end{thm*}
Since in the asymptotic regime, $f_{r}\left(r\right)=\delta\left(r-1\right),$
the conformal mapping becomes\[
\cos\theta=\frac{2X}{1+X^{2}},\,\,\,\sin\theta=\frac{X^{2}-1}{X^{2}+1},
\]
it can be easily verified that\[
\cos\left(\frac{\theta}{2}-\frac{\pi}{4}\right)=\frac{X}{\sqrt{1+X^{2}}}=\tilde{X}
\]
so that a simple change of variable yields\[
\theta\sim\left(1-\sin\theta\right)^{\mu-1}
\]
as obtained in \cite{Comtet}.

The asymptotic distribution of the angular part has thus maximum Tsallis
entropy with parameter $\tilde{q}$ such that \[
\tilde{q}=\frac{\mu-2}{\mu-1}<1.
\]

\section{Conclusion}

We have shown how to use the monotonicity of the Shannon or Tsallis
relative entropies to deduce an $H-$theorem for the Shannon and Tsallis
entropy, first in the general case and then in the case of the Brownian
motion on the hyperbolic plane. Three remarkable results have been
observed in this study: first, the natural drift induced by the negative
constant curvature of the hyperbolic plane transforms the asymptotically
Gaussian of the usual Brownian motion on the plane to the Cauchy distribution,
which belongs to the extended family of Tsallis distributions. Secondly,
the addition of a constant positive external drift transforms this
Cauchy behavior to a non-extensive behavior with non-extensivity parameter
directly related to the value of this drift. This appearance of non-extensive
distributions in the context of an underlying curved space remains
to be linked to physically relevant experiments and data. At last,
the Tsallis distributions with $q>1$ on the Poincaré half upper-plane
realization of the hyperbolic plane transform, via the conformal mapping,
into Tsallis distributions with $q<1$ on the unit disk realization
of the hyperbolic plane.

\section{Annex: proof of Shiino's result}

Following Shiino's notations, we consider\[
D_{q}\left(f_{t}\Vert g_{t}\right)=\int_{\mathbb{R}}f_{t}(x)^{q}g_{t}(x)^{1-q}dx
\]
and omit the time and space variables for readability. 
The time
derivative reads\[
\frac{\partial}{\partial t}D_{q}\left(f\Vert g\right)=q\int f^{q-1}\left(\frac{\partial}{\partial t}f\right)g^{1-q}+\left(1-q\right)\int f^{q}g^{-q}\left(\frac{\partial}{\partial t}g\right).
\]
Since $f$ and $g$ are both solutions of the Fokker-Planck equation
(\ref{eq:FPequation}), we deduce\begin{eqnarray*}
\frac{\partial}{\partial t}D_{q}\left(f\Vert g\right) & = & q\int\left(\frac{f}{g}\right)^{q-1}\left(-\frac{\partial}{\partial x}\left(Kf\right)+\frac{\partial^{2}}{\partial x^{2}}\left(Qf\right)\right)\\
 & + & \left(1-q\right)\int\left(\frac{f}{g}\right)^{q}\left(-\frac{\partial}{\partial x}\left(Kg\right)+\frac{\partial^{2}}{\partial x^{2}}\left(Qg\right)\right)
\end{eqnarray*}
The integrals with the drift function $K\left(x\right)$ can be integrated
by parts, yielding respectively\[
-q\int\left(\frac{f}{g}\right)^{q-1}\frac{\partial}{\partial x}\left(Kf\right)=+q\left(q-1\right)\int\left(\frac{f}{g}\right)^{q-2}\frac{\partial}{\partial x}\left(\frac{f}{g}\right)Kf
\]
and\begin{eqnarray*}
-\left(1-q\right)\int\left(\frac{f}{g}\right)^{q}\frac{\partial}{\partial x}\left(Kg\right) & = & q\left(1-q\right)\int\left(\frac{f}{g}\right)^{q-1}\frac{\partial}{\partial x}\left(\frac{f}{g}\right)Kg\\
 & = & q\left(1-q\right)\int\left(\frac{f}{g}\right)^{q-2}\frac{\partial}{\partial x}\left(\frac{f}{g}\right)Kf
\end{eqnarray*}
so that their sum vanishes.

The integrals with the diffusion function $Q\left(x\right)$ are also
integrated by parts according respectively to\begin{eqnarray*}
q\int\left(\frac{f}{g}\right)^{q-1}\frac{\partial^{2}}{\partial x^{2}}\left(Qf\right) & = & -q\int\frac{\partial}{\partial x}\left(\frac{f}{g}\right)^{q-1}\frac{\partial}{\partial x}\left(Qf\right)\\
 & = & -q\left(q-1\right)\int\left(\frac{f}{g}\right)^{q-2}\frac{\partial}{\partial x}\left(\frac{f}{g}\right)\frac{\partial}{\partial x}\left(Qf\right)\\
 & = & -q\left(q-1\right)\int\left(\frac{f}{g}\right)^{q-1}\frac{\partial}{\partial x}\left(\log\frac{f}{g}\right)\frac{\partial}{\partial x}\left(Qf\right)\\
 & = & -q\left(q-1\right)\int f\left(\frac{f}{g}\right)^{q-1}\frac{\partial}{\partial x}\left(\log\frac{f}{g}\right)\frac{\partial Q}{\partial x}\\
 & - & q\left(q-1\right)\int Q\left(\frac{f}{g}\right)^{q-1}\frac{\partial}{\partial x}\left(\log\frac{f}{g}\right)\frac{\partial f}{\partial x}
\end{eqnarray*}
and\begin{eqnarray*}
\left(1-q\right)\int\left(\frac{f}{g}\right)^{q}\frac{\partial^{2}}{\partial x^{2}}\left(Qg\right) & = & -\left(1-q\right)\int\frac{\partial}{\partial x}\left(\frac{f}{g}\right)^{q}\frac{\partial}{\partial x}\left(Qg\right)\\
 & = & -\left(1-q\right)q\int\left(\frac{f}{g}\right)^{q-1}\frac{\partial}{\partial x}\left(\frac{f}{g}\right)\frac{\partial}{\partial x}\left(Qg\right)\\
 & = & -\left(1-q\right)q\int g\left(\frac{f}{g}\right)^{q-1}\frac{\partial}{\partial x}\left(\frac{f}{g}\right)\frac{\partial Q}{\partial x}\\
 & - & \left(1-q\right)q\int Q\left(\frac{f}{g}\right)^{q-1}\frac{\partial}{\partial x}\left(\frac{f}{g}\right)\frac{\partial g}{\partial x}
\end{eqnarray*}
Their sum consists in an integral with the diffusion function $Q$\[
\left(q-1\right)q\int Q\left(\frac{f}{g}\right)^{q-1}\left(\frac{\partial}{\partial x}\left(\frac{f}{g}\right)\frac{\partial g}{\partial x}-\frac{\partial}{\partial x}\left(\log\frac{f}{g}\right)\frac{\partial f}{\partial x}\right)
\]
and an integral with its derivative\[
q\left(1-q\right)\int\left(\frac{f}{g}\right)^{q-1}\left(f\frac{\partial}{\partial x}\left(\log\frac{f}{g}\right)-g\frac{\partial}{\partial x}\left(\frac{f}{g}\right)\right)\frac{\partial Q}{\partial x}.
\]
The second integral is easily seen to vanish, while the first one
can be simplified to\[
\left(q-1\right)q\int Q\left(\frac{f}{g}\right)^{q-1}\left(\frac{\partial}{\partial x}\left(\frac{f}{g}\right)\frac{\partial g}{\partial x}-\frac{\partial}{\partial x}\left(\log\frac{f}{g}\right)\frac{\partial f}{\partial x}\right)=-q\left(q-1\right)\int Qf\left(\frac{\partial}{\partial x}\log\frac{f}{g}\right)^{2}.
\]
Thus the Tsallis divergence $h_{q}\left(f\Vert g\right)=\frac{1}{q-1}D_{q}\left(f\Vert g\right)$
verifies the stated equality.

\end{document}